\documentclass {aa}
\usepackage{graphicx}
\usepackage{txfonts}
\usepackage[authoryear]{natbib}
\bibpunct{(}{)}{;}{a}{}{,}
\newcommand{\src}  {4U\,1901+03}
\newcommand{\ha}  {H$\alpha$}

\def\simless{\mathbin{\lower 3pt\hbox
     {$\rlap{\raise 5pt\hbox{$\char'074$}}\mathchar"7218$}}}   
\def\simmore{\mathbin{\lower 3pt\hbox
     {$\rlap{\raise 5pt\hbox{$\char'076$}}\mathchar"7218$}}}   

\def\msun{~{\rm M}_\odot}

\usepackage{color}

\begin{document}

   \title{Accretion regimes in the X-ray pulsar 4U\,1901+03}

   \subtitle{}
  \author{
        P. Reig\inst{1,2}
        \and
        F. Milonaki\inst{2}
           }

\authorrunning{Reig et al.}
\titlerunning{Accretion regimes in 4U\,1901+03}

   \offprints{pau@physics.uoc.gr}

   \institute{IESL, Foundation for Reseach and Technology-Hellas, 71110, 
                Heraklion, Greece 
         \and Physics Department, University of Crete, 71003, 
                Heraklion, Greece 
                \email{pau@physics.uoc.gr}
         }

   \date{Received ; accepted}

\abstract
{The source 4U\,1901+03 is a high-mass X-ray pulsar than went into outburst in 2003.
Observation performed with the Rossi X-ray Timing Explorer
showed spectral and timing variability, including the detection of
flares, quasi-periodic oscillations, complex changes in the pulse profiles, and pulse phase
dependent spectral variability.
}
{We re-analysed the data covering the 2003 X-ray outburst and focused on
several aspects of the variability that have not been discussed so far. These are 
the 10 keV feature and the X-ray spectral states and their association
with accretion regimes, including the transit to the propeller state at the
end of the outburst.
}
{We extracted light curves and spectra using data from the Rossi X-ray 
Timing Explorer. Low time resolution light curves were used to create 
hardness-intensity diagrams and study daily changes in flux. High time
resolution light curves were used to create pulse profiles. An average
spectrum per observation allowed us to investigate the evolution of the
spectral parameters with time.
}
{
We find that 4U\,1901+03 went through three accretion regimes over the
course of the X-ray outburst. At the peak of the outburst and for a very
short time, the X-ray flux may have overcome the critical limit that marks
the formation of a radiative shock at a certain distance above the neutron
star surface.  Most of the time, however, the source is in the subcritical
regime. Only at the end of the outburst, when the luminosity decreased
below $\sim 10^{36}\,\, (d/10 \,{\rm kpc})^2$ erg s$^{-1}$, did
the source enter the propeller regime.
Evidence for the existence of these regimes comes from the pulse profiles,
the shape of the hardness-intensity diagram, and the correlation of various
spectral parameters with the flux. The 10 keV feature appears to strongly 
depend on the X-ray flux and on the pulse phase, which opens the 
possibility to interpret this feature as a cyclotron line.
}
{}

\keywords{stars: individual: \src,
 -- X-rays: binaries -- stars: neutron -- stars: binaries close --stars: 
 emission line, Be
               }

   \maketitle

\section{Introduction}

The source \src\ was first detected in X-rays by the {\it Uhuru} mission in 1970-1971.
New {\it Uhuru} observations failed to detect the source a year later
\citep{forman76}. \src\ was not detected again until February 2003, when the
source underwent a giant X-ray outburst that lasted for about five months
and in which the X-ray luminosity changed by almost three orders of
magnitude from $\sim 10^{38}$ erg s$^{-1}$ to $\sim 10^{35}$ erg s$^{-1}$,
assuming a distance of 10 kpc \citep{galloway05}.  Renewed X-ray
activity was detected by {\it MAXI}/GSC \citep{sootome11} and {\it
Fermi}/GBM \citep{jenke11} in 2011. However, the 2011 event was a short
($\sim 1$ month) and weak event with a maximum intensity of $\sim$23 mCrab
in the energy range 4--10 keV on MJD 55920 \citep{sootome11}, more than two
orders of magnitude lower than the peak of the 2003 event.


The source \src\ is believed to be a Be/X-ray binary (BeXB), although no optical
counterpart has been detected so far.  Evidence for this classification
comes exclusively from its X-ray timing properties.  \src\ is a transient
X-ray pulsar with a pulse period of 2.76 seconds \citep{galloway05}. The
orbital period of 22.6 days \citep{galloway05,jenke11} locates \src\ in the
region populated by BeXBs of the $P_{\rm orb}-P_{\rm spin}$ diagram
\citep{corbet86}. A quasi-periodic oscillation (QPO) was detected by
\citet{james11}. The frequency of the QPO ($\sim0.135$ Hz) lies in the
milliHertz range, in agreement with other BeXB pulsars \citep[][and
references therein]{paul11}. The pulse profile is complex and displays
variability over the course of the outburst. At high luminosity, it shows a
double-peak structure that changes into single peak towards the end of the
outburst. These changes were interpreted as a change from an accretion
column dominated by radiative shock above the neutron star surface (fan
beam) into a configuration where the accreted mass is decelerated on to the
surface of the neutron star (pencil beam) \citep{chen08}. Pulse-phase
spectroscopy showed that the main pulse peak has the hardest spectrum,
which is a common property of accreting pulsars \citep{lei09}.

The X-ray spectrum shows significant emission above 10 keV, which would
also agree with an accreting pulsar \citep{molkov03,galloway05}.
However, the flux decreases quickly with energy; there is little emission
above 80 keV, in contrast to many high-mass X-ray pulsars. Attempts to fit
the X-ray spectrum over the entire outburst with a consistent model at high
as well as low luminosity have proved to be difficult. An absorbed
power-law and exponential cutoff  \citep{molkov03,james11} or a model consistent with
thermal Compotonization \citep{galloway05} leave residuals, especially at
around 10 keV, indicating that these models do not completely describe the
source spectrum.

Accreting X-ray pulsars go through different accretion regimes associated
with the X-ray luminosity. Observational evidence for the existence of
these accretion regimes comes from the shape of the pulse profiles
\citep{basko76,parmar89a}, the correlation between the energy of cyclotron
lines and flux \citep{becker12,poutanen13,nishimura14,mushtukov15a}, and
the dependence of the X-ray continuum with luminosity
\citep{reig13,postnov15}.

In this work, we revisit the Rossi X-ray Timing Explorer ({\it RXTE})
observations  and perform a new timing and spectral analysis with
emphasis on the residuals at $\sim$ 10 keV. We study the X-ray variability
of \src\ in the context of accretion regimes. Finally, we
investigate  the X-ray spectrum at very low accretion rates, where an
abrupt decrease in flux is observed, which might be associated with a
transition to the propeller state.  We also perform an optical and
infrared photometric analysis in an attempt to determine its optical
counterpart.

\begin{figure*}
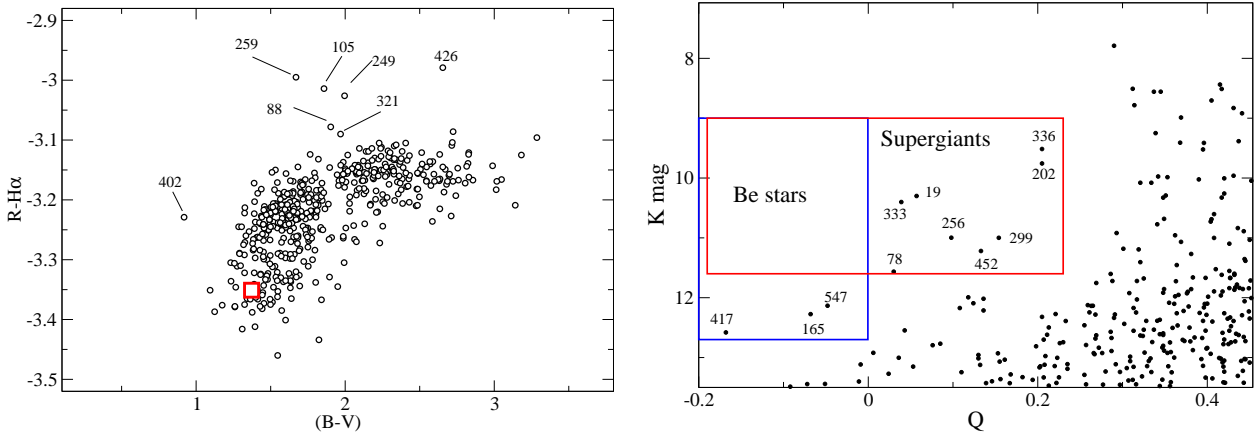

\begin{tabular}{cc}
\includegraphics[width=8cm]{./fig1a.eps} &
\includegraphics[width=8cm]{./fig1b.eps} \\
\end{tabular}
\caption[]{Optical and infrared diagrams used to search
for the optical counterpart. The star represented with a red empty square 
in the left panel corresponds to star 165 from the infrared data. 
The blue and red boxes in the right panel denote the area occupied by 
Be and early-type supergiant stars, respectively.  }
\label{ccd}
\end{figure*}

\section{Observations}

\subsection{X-ray observations}
\label{xobs}

The {\it RXTE} data consist of 64 observation intervals covering the period
JD 2452680.3 (10 February 2003) to JD 2452836.9 (16 July 2003).  {\it RXTE}
carried three instruments: the All Sky Monitor (ASM) 
\citep{levine96}, the Proportional Counter Array \citep[PCA][]{jahoda06},
and the High Energy X-ray Timing Experiment \citep[HEXTE][]{rothschild98}.
In this work, the timing analysis was performed using data from the PCA,
while for the spectral analysis we used both the PCA and HEXTE
data.
 
The PCA covered the energy range 2--60 keV and
consisted of five identical coaligned gas-filled proportional units giving a
total collecting area of 6500 cm2 and providing an energy resolution of 18\%
at 6 keV. The HEXTE was constituted by
two clusters of four NaI/CsI scintillation counters, with a total collecting
area of 2 $\times$ 800 cm2, which are sensitive in the 15--250 keV band with a
nominal energy resolution of 15\% at 60 keV.

Owing to the low Earth orbit of {\it RXTE}, the observations consisted of a number
of contiguous data intervals or pointings interspersed with observational
gaps produced by Earth occultations of the source and passages of the
satellite through the South Atlantic Anomaly. Data taken during satellite
slews, passage through the South Atlantic Anomaly, Earth occultation, and
high voltage breakdown were filtered out.

\subsection{Optical observations and infrared data}

In an attempt to identify the optical counterpart to \src, we obtained
optical photometry in three broad bands, $B$, $V$, and $R$
\citep{bessel90}, and a narrow \ha\ filter. The observations were made with
the 1.3 m telescope of the Skinakas observatory on the night of 22 July
2015 using a 2048$\times$2048 ANDOR CCD with a 13.5 $\mu$m pixel size
(corresponding to 0.28 arcsec on the sky) and thus providing a field of
view of $\sim$9.5 arcmin squared. 

The instrumental magnitudes were used to construct a colour-colour ($R-$\ha\
vs $B-V$) diagram following the procedure described in \citet{reig05a}. Be
stars are expected to occupy the upper (because they are \ha\ emitters) and
left (because they are early-type objects). Additionally, potential candidates
should lie inside the X-ray uncertainty region. 

We also used the 2MASS catalogue \citep{skrutskie06} to compute the
reddening-free quantity $Q=(J-H)-1.70(H-K_S)$ to create a $Q/K_S$ diagram
\citep{negueruela07b}. This diagram is useful to separate early-type from
late-type stars. The majority of stars in Galactic fields are concentrated
around $Q = 0.4-0.5$, corresponding to field K and M stars, while
early-type stars typically have $Q \simless 0$.
We filtered out the infrared data and considered only the best-quality
detections, with signal-to noise ratios higher than 7 and uncertainties of less
than 0.155 mag (Q-flag equal to A or B) and without contamination
or confusion from nearby sources (C-flag equal 0).  Typical errors for
$K_S$ and $Q$ are 0.05 and 0.1 mag., respectively.

\begin{figure}
\center
\includegraphics[width=8cm]{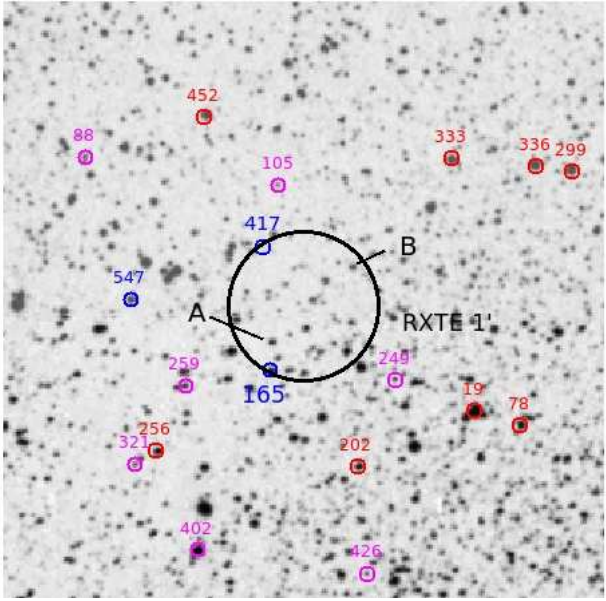} 
\caption[]{Identification of candidates. Red and blue circles 
correspond to possible candidates according to the position in the 
$Q/K_S$ diagram, while purple circles correspond to candidates from the 
optical diagram. The image size is $8.5^{\arcmin} \times 8.5^{\arcmin}$.  
The X-ray position is R.A. = 19h03m37s.1, Decl. = +3 11'31" 
\citep[equinox 2000.0; estimated 1' uncertainty at 90\% confidence][]{galloway03}. 
Stars A and B are not the optical counterpart because they are late-type 
stars \citep{galloway05}. Image taken from the Digitized Sky Survey (DSS),
available through the Skyview Virtual 
Observatory (http://skyview.gsfc.nasa.gov/current/cgi/titlepage.pl)

}
\label{chart}
\end{figure}

Figure~\ref{ccd} shows the optical colour-colour and the infrared $Q/K_S$
diagrams. The blue and red boxes in the right panel of this figure denote
the area where Be and early-type supergiant stars are expected to appear,
respectively \citep{negueruela07b}. These candidates are represented in
Fig.~\ref{chart} by red and blue circles, while the candidates from the
optical colour-colour diagram are indicated by purple circles.  The optical
photometry did not provide any strong candidate. All potential candidates
are far away from the {\em RXTE} X-ray error circle. The infrared
data mark stars 165 and 417 as possible candidates for a
Be star (Fig.~\ref{chart}). However, star 165 (V=15.4 mag) does not
stand out in the optical diagram (red empty square). We note that the
infrared observations were made many years before the optical observations.
Be stars are known to go through active (with a well-developed equatorial
disk) and disk-loss phases on timescales of years. Therefore, star 165
might still be the correct counterpart if it were found to be in a low phase
(without the disk) during the optical observations. Star 417 was too
faint to obtain photometry trough the \ha\ filter.

We obtained a low-resolution (2 \AA/pixel) optical (3500-7300 \AA)
spectrum  of star 165 from the Skinakas observatory on 23 June 2016. Balmer
lines in absorption are the most prominent features.  The presence of weak
HeI lines at 4016 \AA, 4713 \AA, and 4921 \AA\ cannot be ruled out,
although HeI 6678 \AA\ is not seen. Overall, the optical spectrum resembles
that of a late B-type star (later than B5) or early A-type star. The
equivalent width of the \ha\ and $H\beta$ lines (12 \AA\ and 15 \AA,
respectively) would confirm this classification \citep{jaschek87}. The lack
of strong OII lines precludes a supergiant star, although a slightly
evolved star (luminosity class III) cannot be excluded. Although it is
unlikely that star 165 is the optical counterpart, a higher resolution
spectrum with a higher signal-to-noise ratio, especially in the blue part,
is needed to clarify its spectral type.

\section{Results}
\label{results}

The information on the X-ray pulsar \src\ is basically
limited to the {\it RXTE} observations. The analysis of these observations
allowed the  discovery of pulsations and the determination of the orbital
parameters \citep{galloway05}, the study of the pulse profile variability,
both with time and energy \citep{chen08,lei09}, and the detection of a QPO
\citep{james11}. We have preformed a correlated X-ray timing
and  spectral analysis to study whether this source exhibits spectral
states as it evolves through the X-ray outburst. The presence of states has
been associated with different accretion regimes \citep{reig13}.  In
addition to the super- and sub-critical regimes, we also examined whether
\src\ transited to the propeller regime at the end of the outburst.

\subsection{Hardness-intensity diagram}

\begin{figure*}
\resizebox{\hsize}{!}{\includegraphics{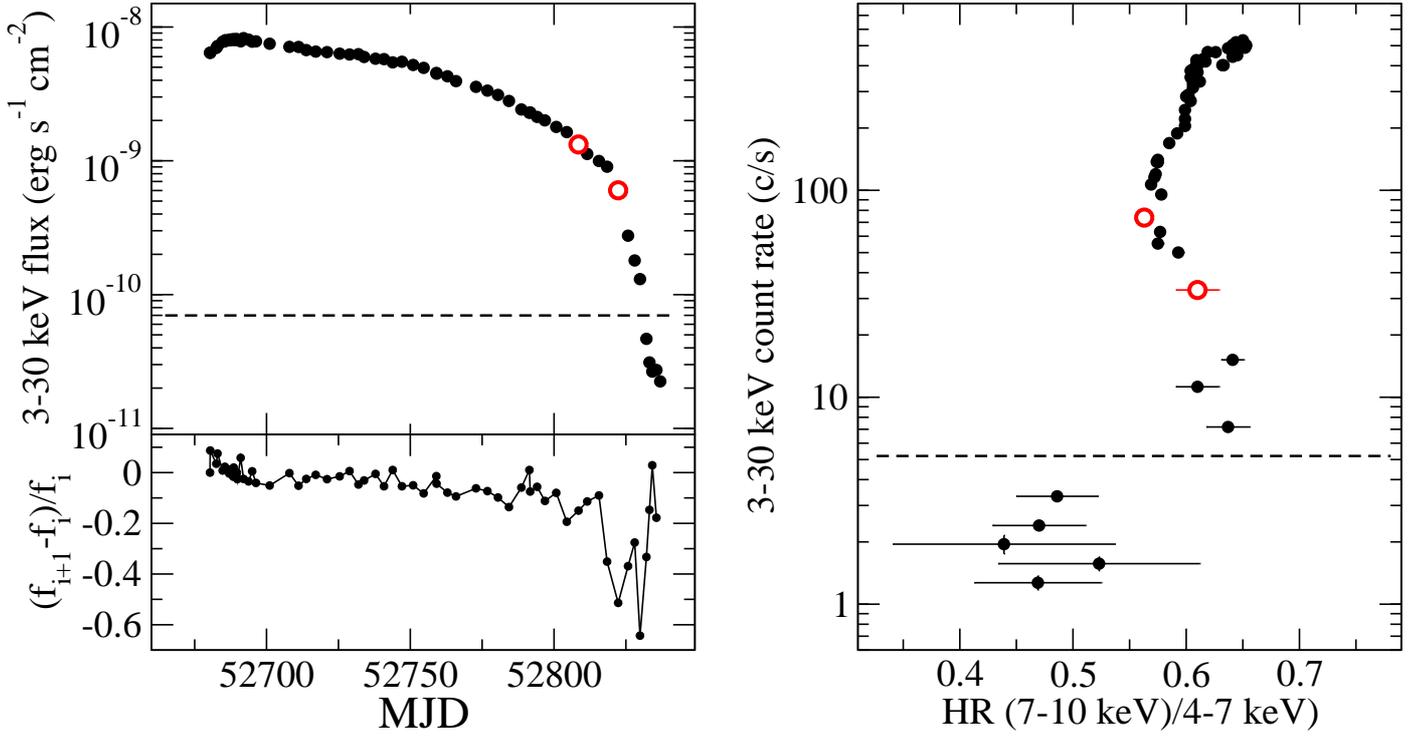}} 
\caption[]{{\em Left}: X-ray flux. Each point represents the average
of one observation.  {\em Right}: Hardness-intensity diagram.  The 
horizontal dashed line marks the lowest flux below which the source is 
expected to enter the propeller regime. Open circles indicate the breaks
in the light curves and HID.}
\label{hid}
\end{figure*}

For each of the individual X-ray observations, we extracted light
curves and spectra. The average intensity of each observation was used to
study the evolution over time and to create
colour-intensity diagrams (HID). A HID is a plot of the X-ray intensity as
a function of hardness. The hardness ratio is the ratio between
the photon counts in two broad energy bands. Figure~\ref{hid} shows the light
curve of the outburst and the HID constructed with a hardness ratio
defined by the 4--7 keV and 7--10 keV energy bands.

The first PCA observation was made $\sim$10 days before the source reached
the peak of the outburst. During this short rise, the source moved toward
the right (i.e. hardened) in the HID. As the count rate decreased following
the decay of the outburst, the spectrum became softer. A deviation from a
smooth decay is recorded at around $\sim$70 c s$^{-1}$, in which the source
suddenly became harder  (Fig.~\ref{hid}). The lowest flux points correspond
to the softest spectrum.  

\subsection{X-ray spectra}
\label{xrayspec}

An energy spectrum was obtained for each pointing. We used Standard 2 mode
data from the PCA (PCU2 only) and Standard (archive) mode from the HEXTE
Cluster A, with a time resolution of 16 s. The PCA and HEXTE spectra were
extracted, background-subtracted, and dead-time corrected. For the PCA, the
3--30 keV energy range was retained, while the HEXTE provided a partially
overlapping extension from 25 to 100 keV. A systematic error of 0.6\%
was added quadratically to the statistical one.  The two resulting spectra
were simultaneously fitted with XSPEC v. 12.8.0 \citep{arnaud96}.

The X-ray spectra were fitted with a multi-component model consisting of an
absorbed ({\tt PHABS} in XSPEC terminology) power law and exponential
cutoff ({\tt CUTOFFPL} ) plus two discrete components: an emission line at
6.4 keV representing fluorescent iron line emission and a broad absorption
component around 10 keV. These two components were modelled with Gaussian
profiles ({\tt GAUSS} and {\tt GABS}, respectively). Therefore the model
consisted of the following spectral parameters: hydrogen column density,
$N_H$, exponent of the power law or photon index, $\Gamma$, cut-off energy,
$E_{\rm cut}$, and the line energy and width of the two Gaussians, $E_{\rm
Fe}$, $E_{\rm abs}$, $\sigma_{\rm Fe}$, and $\sigma_{\rm abs}$. In addition,
each of these components has a normalisation coefficient. For the Gaussian lines, the normalisation (or line depth) is related to the
optical depth, which at the line centre is $\tau_{\rm c}={\rm
norm}/(\sqrt{2\pi}\,\sigma_{\rm abs})$. The photon distribution of each
model component is given in Table~\ref{models}. To account for
inter-calibration uncertainties between the two instruments, a constant
factor was introduced. This factor was fixed to one for the PCA data and
allowed to vary freely for the HEXTE data.   At low flux
($\sim 5\times 10^{-10}$ erg cm$^{-2}$ s$^{-1}$ or $L_X\sim 6\times
10^{36}$ erg s$^{-1}$, assuming a distance of 10 kpc), the signal-to-noise
ratio of the continuum deteriorates and the spectrum is consistent with a simple
power law, where neither the cut-off nor the broad Gaussian components are
required to obtain an acceptable fit.

Soft X-rays are absorbed by the interstellar medium through the
photoelectric effect \citep{morrison83,balucinska92}.  Because of the steep
energy dependence of the cross-section with the abundances assumed for the
Galaxy \citep{anders89}, photoelectric absorption is significant at
energies below $\sim$ 1 keV. Initially, the hydrogen column density was let
free to vary. Its value remained roughly constant at a value of $3.5\times
10^{22}$ cm$^{-2}$ during most of the outburst and displayed a small
decrease towards low-flux observations. Because the trend is not
statistically significant and the PCA instrument cannot constrain this
parameter well (it is sensitive for energies above 3 keV), we set
$N_H=3.3\times 10^{22}$ cm$^{-2}$ and kept it fixed during the fitting
procedure.

\begin{figure}
\includegraphics[width=8cm]{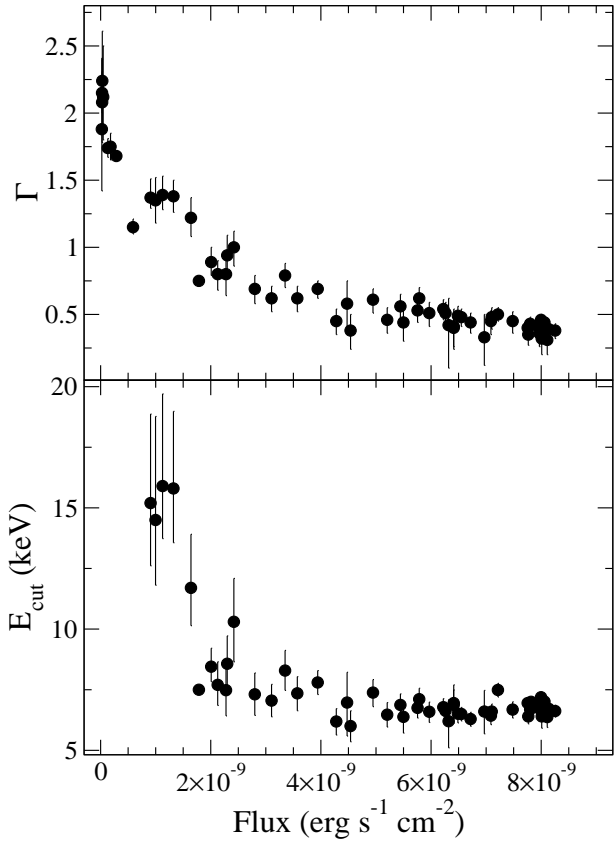} 
\caption[]{Evolution of the photon index and cutoff energy as a function of
 the 3--30 keV flux.}
\label{specpar}
\end{figure}

The photon index and cut-off energy anti-correlate with the X-ray flux
(Fig.~\ref{specpar}). The hardest spectrum is seen at the peak of the
outburst. 

\begin{figure}
\includegraphics[width=8cm]{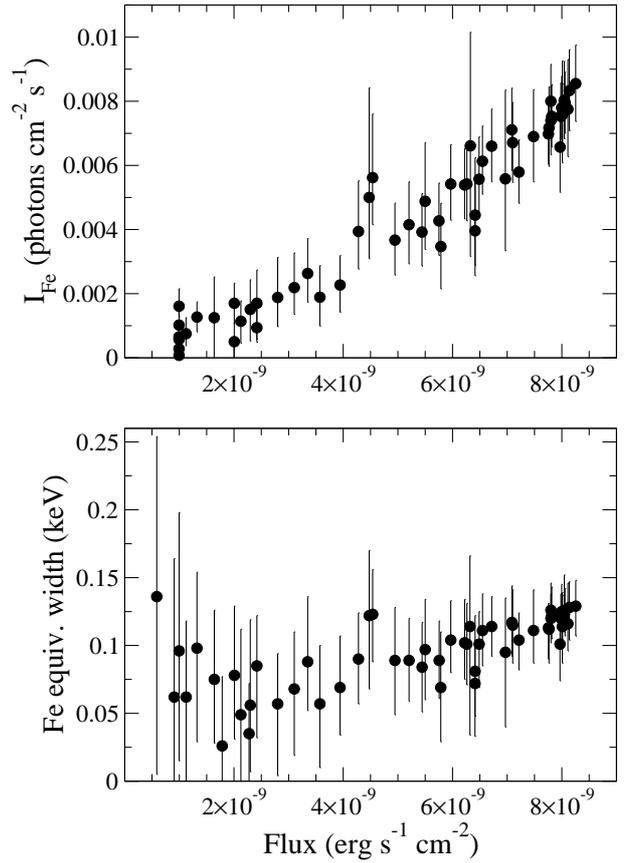} 
\caption[]{Iron line flux and equivalent width as a function of continuum
the 3--30 keV X-ray flux.}
\label{ironline}
\end{figure}

Iron line emission is commonly observed in the X-ray spectra of accreting
X-ray pulsars \citep{torrejon10,gimenez15}. The most likely origin is Fe K
line fluorescence that is produced as a consequence of the X-ray
illumination of matter. The result that we find for \src\ agrees with those
observed in other BeXBs \citep{reig13}: {\em i)} the energy  of the iron
line did not change significantly and is consistent with cold non-ionised
iron. The average and standard deviation of the line energy is $E_{\rm
Fe}=6.5\pm0.1$ keV, {\em ii)} the flux of the line  is well correlated with
the flux of the continuum, and {\em iii)} the equivalent width is
insensitive to luminosity changes (Fig.~\ref{ironline}). The line width
close to the peak of the outburst appears to be somewhat broader
$\sigma=0.8\pm0.3$ keV. Below $8\times 10^{-9}$ erg cm$^{-2}$ s$^{-1}$ the
width is consistent with a narrow line ($\sigma < 0.5$ keV).  However,
given the modest spectral resolution of the PCA, good fits were obtained by
fixing the width at $\sigma_{\rm Fe}=0.5$ keV in all observations. To
better constrain the spectral parameters of the power law, cut-of,f and
absorption feature, we fixed the energy of the iron line and width to 6.5
keV and 0.5 keV, respectively,  during the extraction of the spectral
continuum parameters and the estimation of their errors.

\begin{figure}
\resizebox{\hsize}{!}{\includegraphics{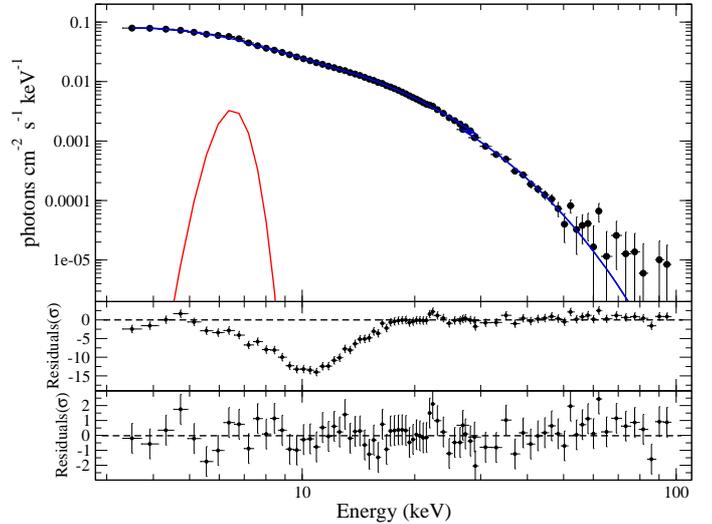}} 
\caption[]{X-ray spectrum of \src\  on 2003-02-12 (OBSID
70096-01-01-03). The middle panel shows the residual when
the {\tt GABS} component is not included in the fit. The bottom panel shows
the residuals for the best-fit model. }
\label{spec}
\end{figure}

\begin{table}
\caption{Photon distributions of the phenomenological models used in the spectral analysis. }
\label{models}      
\centering          
\begin{tabular}{ll}
\hline\hline
XSPEC model     &Photon energy distribution              \\
\hline
{\tt PHABS}     &$A(E)=e^{-N_{\rm H} \sigma_{MM}(E)}$ \\
{\tt CUTOFFPL}  &$C(E)=K\,E^{-\Gamma} e^{-E/E_{\rm cut}}$       \\
{\tt POWER}     &$P(E)=K\,E^{-\Gamma}$  \\
{\tt HIGHECUT}  &$H(E)= 1 \,\,E \leqslant E_{\rm cut} ; H(E)=e^{(E_{\rm cut}-E)/E_{\rm fold}} \,\,E \geqslant E_{\rm cut}$      \\
{\tt GAUSS}     &$G(E)= K_1 \frac{1}{\sigma \sqrt{2\pi}} e^{-0.5\left(\frac{E-E_l}{\sigma}\right)^2}$\\
{\tt GABS}      &$G_{\rm abs}(E)=e^{\left(-K_2 \frac{1}{\sigma_0 \sqrt{2\pi}}\,e^{-0.5\left(\frac{E-E_0}{\sigma_0}\right)^2}\right)}$   \\
{\tt EDGE}      &$M(E)=1 \,\,E \leqslant E_c ; M(E)=e^{-D(E/E_c)^{-3}} \,\,E \geqslant E_c$\\
{\tt COMPTT}    &see \citet{titarchuk94,hua95} \\
{\tt COMPMAG}   &see \citet{farinelli12} \\
\hline
\multicolumn{2}{l}{$\sigma_{MM}(E)$: 0.03-10 keV interstellar photoelectric absorption} \\
\multicolumn{2}{l}{\hspace{1.1cm} cross-section \citep{morrison83}.} \\
\multicolumn{2}{l}{$N_{\rm H}$: equivalent hydrogen column.} \\
\multicolumn{2}{l}{K: normalisation in photons keV$^{-1}$ cm$^{-2}$
s$^{-1}$ at 1 keV.} \\
\multicolumn{2}{l}{$\Gamma$: power-law photon index.}   \\
\multicolumn{2}{l}{$E_{\rm cut}$: cut-off energy in keV.}       \\
\multicolumn{2}{l}{$E_{\rm fold}$: folding energy in keV.}      \\
\multicolumn{2}{l}{$E_l$: iron line energy in keV.}     \\
\multicolumn{2}{l}{$\sigma$: iron line width in keV.}   \\
\multicolumn{2}{l}{$K_1$: total photons cm$^{-2}$ s$^{-1}$ in the line.} \\
\multicolumn{2}{l}{$E_{0}$: central energy of the absorption line in keV.}   \\
\multicolumn{2}{l}{$\sigma_0$: width of the absorption line in keV.}   \\
\multicolumn{2}{l}{$\tau_0=K_2/(\sqrt{2\pi}\,\sigma_0)$, optical depth of the absorption line.} \\
\multicolumn{2}{l}{$E_c$: threshold energy} \\
\multicolumn{2}{l}{$D$: absorption depth at the threshold} \\
\end{tabular}
\end{table}

\subsection{10 keV feature}

\begin{figure}
\resizebox{\hsize}{!}{\includegraphics{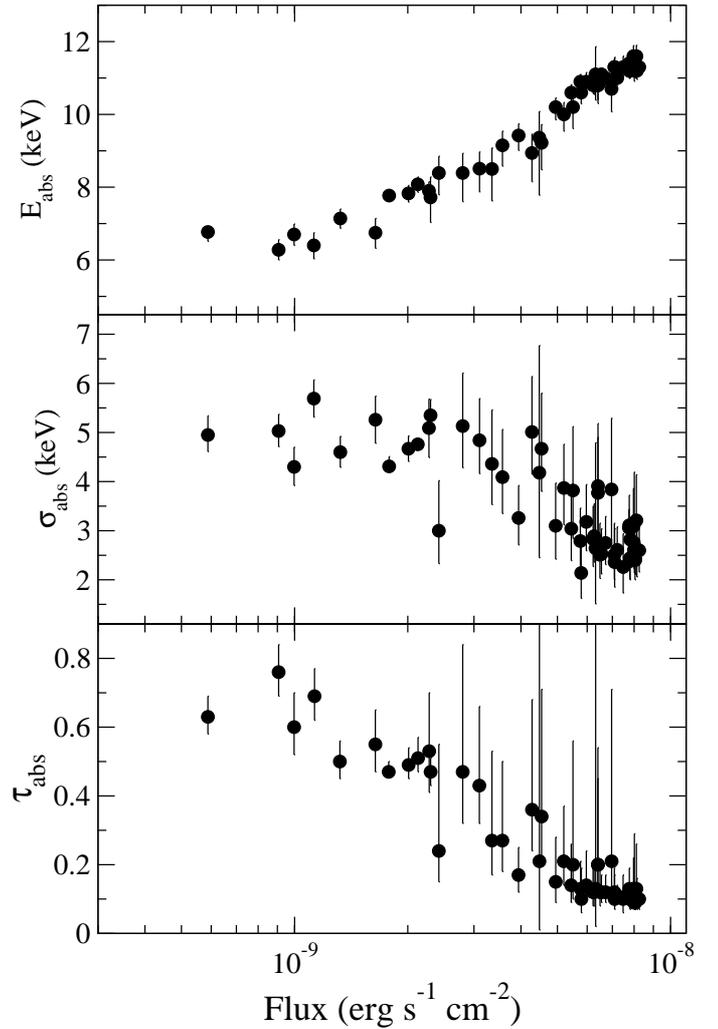}} 
\caption[]{Evolution of the 10 keV feature parameters with X-ray flux (3--30 keV).}
\label{gabs}
\end{figure}

The 10 keV feature is a component that we found in all the spectra of \src\ with an X-ray flux
above $\sim 5\times 10^{-10}$ erg cm$^{-2}$ s$^{-1}$ ($L_X\sim 6\times
10^{36}$ erg s$^{-1}$, assuming a distance of 10 kpc). Without this
component, the fits were unacceptable (see Fig~\ref{spec}). Below this
limit the spectra are too noisy and the component is not statistically
significant.  Its central energy increases as the flux increases
(Fig.~\ref{gabs}). This clear correlation and the relatively large
amplitude change (a factor of two) in $E_{\rm abs}$ motivated us to
investigate further whether the absorption feature that we observed in
\src\  can be attributed to a cyclotron resonant scattering feature (CRSF). We
followed three lines of action: {\em i)} we checked whether the 10 keV
feature in \src\ is  an instrumental effect, {\em ii)} we investigated whether
it might result from improper continuum fitting, and {\em iii)} we performed
pulse-resolved spectroscopy.   

One way to discard an instrumental origin for this feature is to compare
the spectrum of \src\ with that of the Crab nebula. The Crab
nebula is a well-known
calibration source with a pure power-law spectrum, without any other
features \citep{weisskopf10}, at least in the PCA energy band. We selected
the Crab nebula observation made on 13 February 2003 (OBSID 70802-01-08-00) to be
close to the observations of \src. We extracted the X-ray
spectrum using the same configuration as the source (see
Sect.~\ref{results}). The best fit ($\chi^2=50.6$ for 52 degrees of
freedom)  was achieved with an absorbed power law with a hydrogen column
density $N_H=(0.4\pm0.1)\times  10^{22}$ cm$^{-2}$, photon index
$\Gamma=2.104\pm0.007$, and normalisation $10.6\pm0.2$ photons keV$^{-1}$cm
s$^{-1}$ at 1 keV. No residuals are apparent at or around 10 keV. Thus we
conclude that the deficit of photons at around 10 keV is not an
instrumental effect.

To check whether the residuals around 10 keV could be caused by improper
model fitting of the continuum, we tried different combinations of model
components. The photon distribution of the various phenomenological models
are given in Table~\ref{models}. We also tried more physical models that
explain the X-ray continuum of pulsars as inverse Comptonization of
low-energy photons. Here we used  the model developed by \citet{hua95},
which describes Comptonization of soft photons in a hot plasma (named {\tt
COMPTT} in XSPEC), and which is the one used by \citet{galloway05}, and
that of \citet{farinelli12}, which describes the spectral formation in the
accretion column onto the polar cap of a magnetized neutron star, with both
thermal and bulk Comptonization processes taken into account ({\tt COMPMAG}
in XSPEC).  When phenomenological models were used, the spectra were fitted
with the function  $f(E)={\tt PHABS}*({\rm COMP1}+{\tt GAUSS})*{\rm
COMP2}$, where COMP1 is  {\tt CUTOFFPL} or {\tt POWER*HIGHECUT} and COMP2
{\tt GABS} or {\tt EDGE}. When Comptonization models were used, the spectra
were fitted with the function  $f(E)={\tt PHABS}*({\rm COMPTOM}+{\tt
GAUSS})$, where  COMPTOM refers to one of the two Comptonization models
shown in Table~\ref{models}. For this analysis, we used the observation
made on February 19, 2003 (obsid 70068-22-01-04), but other high or
moderate flux observations produced the same result. We found the following
results:

\begin{itemize}

\item[-] The replacement of {\tt CUTOFFPL} by {\tt POWER*HIGHECUT} did not
produce any significant change in the residuals.  Both models give
$\chi^2=42$ for 45 degrees of freedom (dof).

\item[-] The replacement of the {\tt GABS} component by an absorption edge
({\tt EDGE}) reduced but did not eliminate the residuals at $\sim$10 keV.
The final fit was not acceptable with $\chi^2=115$ for 46 dof.

\item[-]  With the use of Comptonization models, the 10-keV absorption
feature becomes insignificant {\em } when the parameters (line and width) of
the emission line component are let free ($\chi^2=36$ for 47 dof).
However, in this case, the best-fit line parameters are unrealistic ($E_{\rm
Fe}=6.1-6.2$ keV,  $\sigma_{\rm Fe}\simmore$ 1.5 keV) for a component
representing fluorescent iron line emission \citep{gimenez15}. In this
case,  the Gaussian component tries to fit the residuals at $\sim$10 keV,
mimicking the bump model of \citet{klochkov07}. Fixing the line energy
and width to reasonable values, $E_{\rm Fe}=6.4-6.6$ keV and $\sigma_{\rm
Fe}\simless 0.5$ keV, the Comptonization models cannot account for the
deviation from the fit in the range 6-12 keV. The fit can be improved
by adding a blackbody component with $kT\approx 1$  keV, that is, without the
need for a specific component related to the deficit of photons at 8--10
keV. This is the model used by \citet{galloway05}. However, even this
multi-component model cannot entirely remove the residuals at energies below 10 keV. Reduced $chi^2\sim 1$ can be obtained provided a larger
systematic error of 1\% is used , see \citet{galloway05}.

\end{itemize}

\begin{figure}
\resizebox{\hsize}{!}{\includegraphics{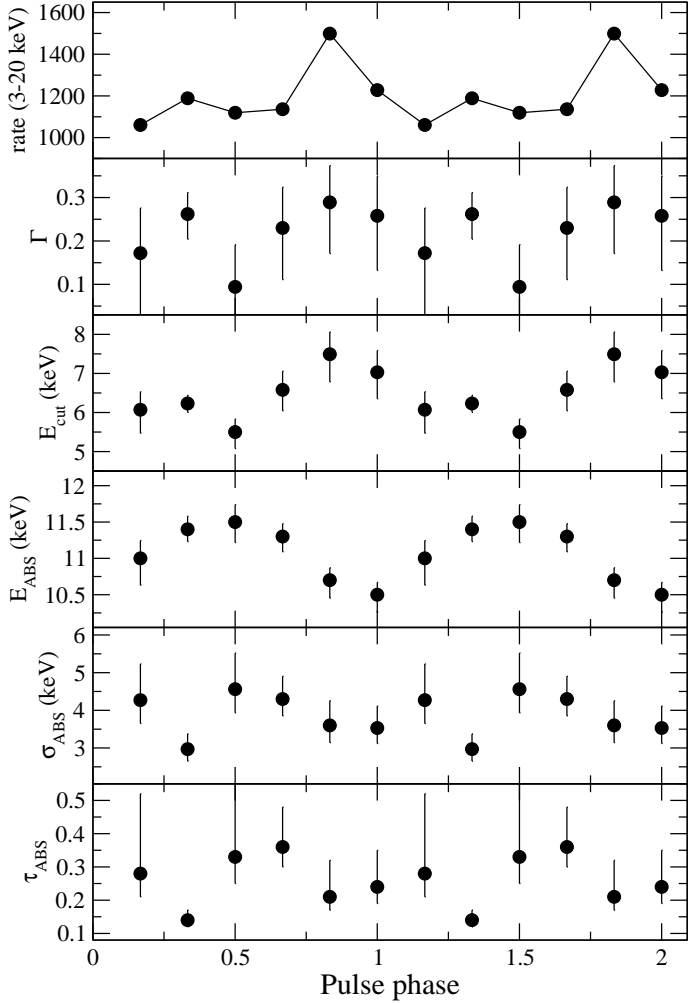}} 
\caption[]{Variability of the spectral parameters with 
pulse phase for observation 70096-01-01-03. Shown are
$1\sigma$ errors.}
\label{pps}
\end{figure}

If the 10 keV feature had a magnetic origin, we would expect to observe
variability of the component parameters with pulse phase \citep{heindl04}.
We performed a pulse phase resolved spectroscopic analysis using
observation 70096-01-01-03 (2003-02-12).  This observation was chosen
because with a total exposure time of 5.1 ks and relatively high intensity,
it represents a good compromise between a good signal-to-noise ratio and a high
number of pulse cycles. But it also avoids contamination between
neighbouring phase bins owing to the uncertainty in the  spin
period that a longer observations might introduce.  First, the pulse
period has to be determined. To this end, we extracted a light curve with a
time resolution of $2^{-8}$ s using {\it GoodXenon} event data. The light
curve was divided into 256-s segments, and for each segment we obtained the
pulse profile by folding the light curve on a trial period. Taking the first
profile as reference, we cross-correlated each profile with the reference
profile. The slope of the linear fit of the time shifts with respect to the
mid-time of the segments provided by the correction for the trial period. We
repeated the procedure until the data points did not show any trend. The
final pulse period was $2.762410\pm0.000031$ s. The pulse profile was
divided into six equally spaced bins to provide six 2.9--20 keV  spectra.
Each spectrum was fitted with the same model as the phase-average spectra
of Sect.~\ref{xrayspec}. We fixed the hydrogen column density
and the iron line
and width to their average values, namely, $N_H=3.3\times 10^{22}$
cm$^{-2}$, $E_{\rm Fe}=6.5$ keV, and $\sigma_{\rm Fe}=0.5$ keV. A  phase
modulation is clearly seen in the centroid energy of the 10 keV
feature (Fig.~\ref{pps}). 

\section{Discussion}

\subsection{Spectral states}

Be/X-ray pulsars that go through large amplitude changes in luminosity
exhibit spectral states when the source luminosity is significantly higher than
the critical luminosity \citep{reig13}. The critical luminosity  indicates
whether the radiation pressure of the emitting plasma is capable of
decelerating the accretion flow \citep{basko76,becker12,mushtukov15b}. The
critical luminosity defines two accretion regimes. If $L_X > L_{\rm crit}$
(super-critical regime), then radiation pressure is so high that is capable
of stopping the accreting matter at a certain distance above the neutron star
surface, which formes a radiation-dominated shock. If $L_X < L_{\rm crit}$
(sub-critical regime), then the accreting material reaches the neutron star
surface, heating it through Coulomb collisions with thermal electrons or through
nuclear collisions with atmospheric protons \citep{harding94}. 

Super-critical sources trace two distinct branches in their HID that were
called horizontal and diagonal branches \citep{reig08,reig13}.  The
horizontal branch corresponds to a low-intensity state that the source
traces during the beginning and end of the outburst, and it is characterised
by fast colour and spectral changes and high X-ray variability. The
diagonal branch is a high-intensity state that emerges when the X-ray
luminosity exceeds the critical limit.  To clearly distinguish two branches
in the HID, $L_{\rm peak}/L_{\rm crit} >> 1$ and the source luminosity must
remain above the critical luminosity long enough so that the observations
can sample this high state. \citet{reig13} also found that the photon index
anti-correlates with X-ray flux in the horizontal branch, but correlates with
it in the diagonal branch. Thus the critical luminosity may be estimated by
measuring the flux at the break that marks the transition between branches
in the HID or at the point where the slope changes in the X-ray flux versus
photon index diagram.

In contrast to 4U\,0115+63, EXO\, 2030+375, V 0332+53 or KS\, 1947+300
\citep[see Fig. 4 in][]{reig08} and \citep[Fig. 6 in][]{reig13}, the HID of
\src\ does not show the two distinct branches (Fig.~\ref{hid}), nor does the
photon index show a clear change from negative to positive correlation
with flux (Fig.~\ref{specpar}). This indicates that \src\  remained  in the sub-critical regime most of
the time. The variability in the pulse profiles
confirms this result. In the super-critical regime, X-ray photons are
expected to escape the accretion column from the sides, that
is, perpendicular
to the magnetic field, radiating in a fan beam. In the sub-critical
regime, the photons escape parallel to the magnetic field in a pencil
beam.  These types of complex changes in the pulse profile with luminosity
have been seen in other X-ray pulsars
\citep{parmar89a,mukerjee00,sasaki12,malacaria15}.  In \src, a double-peak
pulse profile is observed. However,  the relative strength of the two
pulses of the double-pulse profile at low flux reversed compared to the
double peak structure at high flux \citep{chen08}. This result was
interpreted in terms of a luminosity-dependent emission profile of the
pulsar, where the emission pattern changes from being dominated by the
fan beam at high flux to correspond to a pencil beam at low flux
\citep{mukerjee00,chen08}. The mixed contribution of
fan and pencil beams even at the highest luminosity \citep{chen08} implies
that the source was not in a pure super-critical regime, which would be
characterised a a fan beam only \citep[see Fig. 1 in][]{becker12}.

On the other hand, a flattening toward the peak of the outburst is observed
in the $F_X-\Gamma$ relation (see Fig.~\ref{specpar}), which might indicate
that the source reached the critical luminosity close to the peak of the
outburst. Both the HID and the $F_X-\Gamma$ diagram resemble that of  XTE\,
J0658--073 \citep[see Fig. 3 in][]{reig13}. The similarity between \src\
and  XTE\, J0658--073 is reinforced by the fact that the two sources
display X-ray flaring behaviour during the peak of the outburst
\citep{james11,nespoli12}. For XTE\, J0658--073, \citet{reig13} argued that
$L_{\rm peak}/L_{\rm crit} \simmore 1$ but that the source did not remain in the super-critical regime for a substantial period of time, hence it
could not develop a clear diagonal branch. The same situation is likely to
have occurred in \src.

\subsection{10 keV feature}

A peculiar feature in the X-ray energy spectra of \src\ is an
absorption-line-like profile whose energy varies in the range 6--12 keV
(Fig.~\ref{spec}). After fitting the X-ray spectra with a power law, many
X-ray pulsars leave significant residuals  around 10 keV. This feature is
generally referred to as the 10 keV feature \citep{coburn02}. Although it
appears to be a common feature in many X-ray pulsars, its origin and nature
are not known. An instrumental origin is ruled out because it has been 
observed with different instruments onboard {\it RXTE}, {\it Ginga} and
{\it BeppoSAX} \citep[][and references therein]{coburn02}. Because
there is no distinct variability pattern, it is difficult to
interpret. We summarise the most relevant characteristics of this feature below.

\begin{itemize}

\item[-] It may be modelled as an emission feature as in 4U\,0115+63
\citep{ferrigno09,muller13} and EXO\,2030+375 \citep{klochkov07}, or as an
absorption feature as in XTE J0658-073 \citep{mcbride06,nespoli12} or
XTE\,J1946+274 \citep{muller12}. In either case, a Gaussian profile is used
(see Table~\ref{models} for the functional form of these components). 

\item[-] The feature has been observed  regardless of whether a cyclotron
line is present or not. This led \citet{coburn02} to conclude that the
component probably is not a magnetic effect. It may easily be mistaken as a
cyclotron line. For example, in EXO 2030+375, this additional broad emission
component is not needed when two absorption lines at $\sim$10 keV and
$\sim$20 keV are used \citep{klochkov07}, which could be interpreted as
cyclotron lines \citep{wilson08}.

\item[-] The central energy of the feature does not generally vary in time
or flux \citep{nespoli12,muller12}, although a weak anti-correlation with
flux was reported by \citet{muller13} in 4U\,0115+63.

\end{itemize}

Theory predicts that the energy of the cyclotron line observed in many
accreting pulsars shows a positive correlation with flux when the pulsar is
in the sub-critical regime and a negative correlation in the super-critical
regime \citep{becker12,nishimura14,mushtukov15a}. In the previous section
we argued that the source is in the sub-critical regime most of
the time.  The positive correlation of the energy of the absorption feature
with flux that we measure in \src\ (Fig.~\ref{gabs}) would support the
association of the 10 keV feature with a cyclotron line. The width and
depth of the feature are also typical of accreting pulsars
\citep{coburn02}. The shape of cyclotron lines in
accreting X-ray pulsars depends strongly on the viewing angle, hence on the
pulse phase \citep{isenberg98,araya00,heindl04}. Additional support for the
interpretation of the 10 keV feature as a cyclotron line therefore comes from the
variation of the feature parameters with pulse phase (Fig.~\ref{pps}).

On the other hand, if this feature were a cyclotron line, then the
critical luminosity would be of the order of $10^{37}$ erg s$^{-1}$
\citep{becker12,mushtukov15b}. Thus not only $L_{\rm peak}/L_{\rm crit}
\sim 8$ (assuming a distance of 10 kpc), but also $L_X/L_{\rm crit}> 1$ for
most part of the outburst, and a distinct diagonal branch should be observed
in the HID.  We note, however, that this conclusion is based on a rather
uncertain distance. A distance larger than $\sim 8$ kpc would argue against
the 10 keV feature as a cyclotron line because $L_X/L_{\rm crit} > 1$. To
have $L_{\rm peak}/L_{\rm crit} \sim 1$, the distance to the source should
not be larger than $\sim$4 kpc. At this distance, an early-type B star
should be easily detectable with photometric observations, unless the
optical extinction is very high, $A_V\simmore 10$ mag. The relatively high
column density of the spectral fits supports high extinction toward the
source. 

In summary, the inclusion of an absorption  component to account for the
residuals at 8--10 keV provides acceptable fits even with a small
systematic error of 0.6\%. The variability of the energy of this component
with flux and with pulse phase and the persistence of the residuals
regardless of the continuum model used is reminiscent of a CRSF. However,
the lack of a reliable distance estimate prevents us from establishing a
connection between this feature and accretion models.

\begin{figure}
\resizebox{\hsize}{!}{\includegraphics{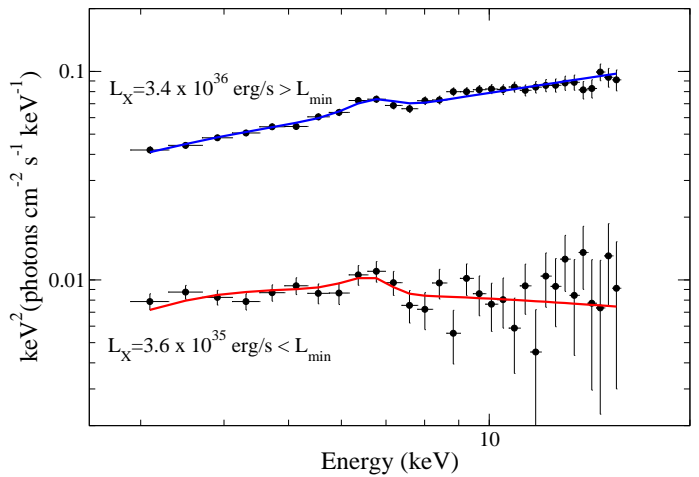}} 
\caption[]{X-ray spectrum of \src\ in the accretor (top) and
propeller (bottom) states.}
\label{propeller-spec}
\end{figure}

\begin{figure}
\resizebox{\hsize}{!}{\includegraphics{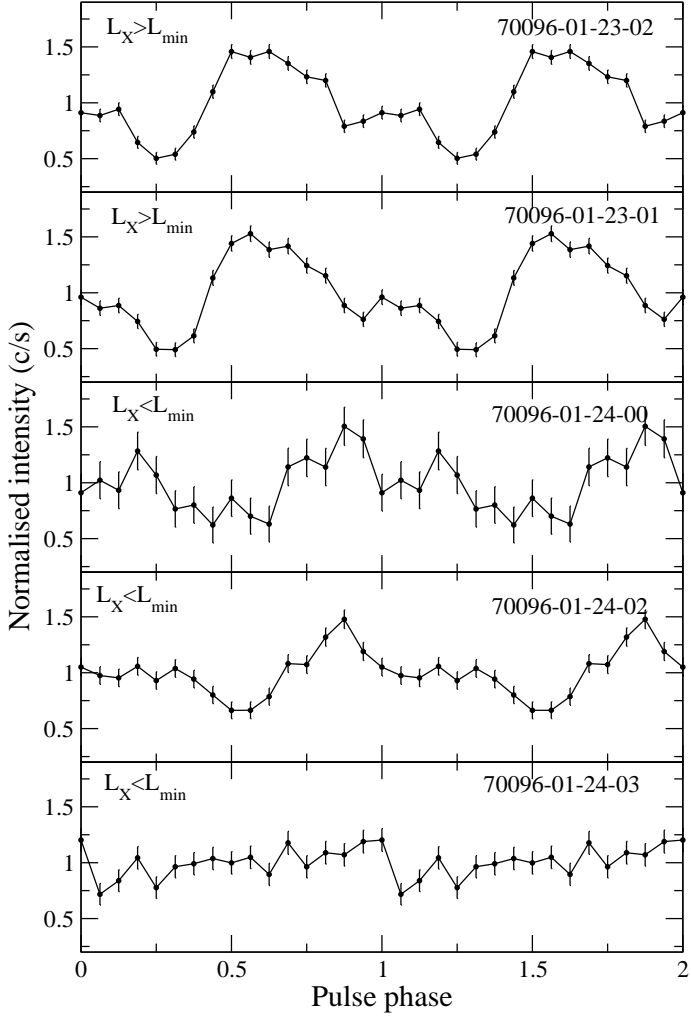}} 
\caption[]{Pulse profiles as a function of luminosity. Below $L_{\rm min}$
the X-ray pulsations vanish.}
\label{propeller-prof}
\end{figure}
\begin{figure}
\resizebox{\hsize}{!}{\includegraphics{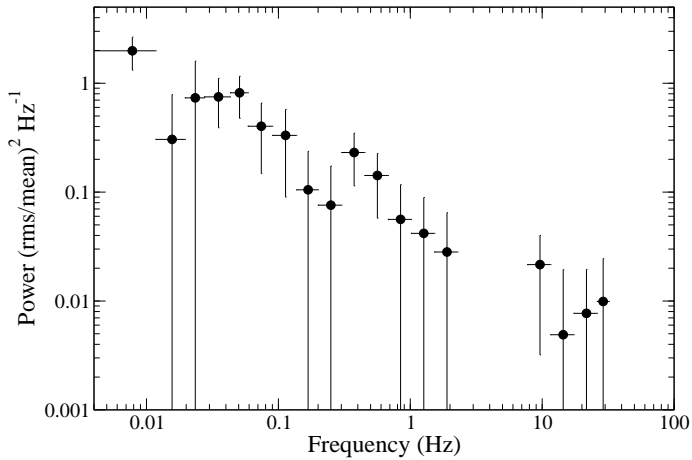}} 
\caption[]{Power spectrum of \src\ in the propeller state in the energy
range 3--15 keV.}
\label{propeller-psd}
\end{figure}

\subsection{Transition to the propeller regime}
\label{prop}

Figure~\ref{hid} shows the light curve of \src. The duration of the decay
phase of the outbursts (from peak to last observation) was $\sim 150$ days.
The source displays several sudden decreases in flux over the course of the
decay.  These abrupt changes in luminosity are apparent in the HID, where a
clear change in the slope of the curve is seen, implying a major change in
the physical conditions of the emission region.

The most significant changes in terms of the fractional amplitude of flux
$(f_{i+1}-f_i)/f_i$) occurred towards the end of the outburst. In
particular, the largest change took place between MJD 52830-58832 (bottom
panel in Fig.~\ref{hid}). These abrupt changes in luminosity have been
attributed to a change in the accretion regime whereby the source transits
from an accretor state to the propeller state \citep{Illarionov75,stella86},
which has been observed in a number of pulsars
\citep{cui97,campana01,campana02,tsygankov16}. 

In the process of mass accretion, the ram pressure of the flow is exerted
on the magnetosphere and is balanced by the magnetic pressure. Therefore,
the size of the magnetosphere is determined not only by field strength but
also by mass accretion rate. In a bright state, the accretion rate is high,
so that the magnetosphere is usually small compared to the co-rotation radius at
which the angular velocity of Keplerian motion is equal to that of the
neutron star. Material is continuously channelled to magnetic poles, and the
X-ray emission from hot spots pulsates. As the accretion rate decreases,
the ram pressure decreases, and thus the magnetosphere expands. As the
magnetosphere grows beyond the co-rotation radius, the centrifugal force
prevents material from entering the magnetosphere, and thus accretion onto
magnetic poles ceases. This is commonly known as the propeller effect 
\citep{Illarionov75,stella86}.

There exists a lowest X-ray luminosity, determined by the lowest
accretion rate, below which the propeller effect sets
in \citep[see e.g.][]{campana02},

\begin{equation}
\label{limlum}
L_{\rm min}(R_{\rm NS}) = 3.9 \times 10^{37} 
k^{7/2} B_{12}^2 P_{\rm s}^{-7/3} M_{1.4}^{-2/3} R_{6}^5 \, \,  {\rm erg \,
s^{-1}}
,\end{equation}

\noindent where $k$ is a constant that accounts for the geometry of the
flow. $k\approx1$ in case of spherical accretion and $k\approx 0.5$ in case
of disk accretion.  $B_{12}$ is the magnetic field strength in units of
$10^{12}$ G, $P_{\rm s}$ the spin period in seconds, and $M_{1.4}$ and
$R_{6}$ the mass and radius of the neutron star in units of $1.4 \msun$ and
$10^6$ cm, respectively. Below this limiting luminosity, the accreting
matter can no longer reach the neutron star surface because it is spun away
by the fast rotation of the magnetosphere. For rapid-rotating pulsars
($P_{\rm spin}\sim 1$ s), $L_{\rm min}(R_{\rm NS})$ is relatively high,
$\sim 10^{36}$ erg s$^{-1}$. For long-pulsing systems ($P_{\rm spin}\sim
100$ s), $L_{\rm min}(R_{\rm NS}) \sim 10^{32}-10^{33}$ erg s$^{-1}$
\citep{stella86,campana01,reig14c,tsygankov16}. Assuming the typical mass
and radius of a neutron star, $k=1$, and a magnetic field of $\sim (3-5)
\times 10^{11}$ G \citep{galloway05,james11}, the lowest luminosity in
\src\ is $L_{\rm min}\sim (4-9) \times 10^{35}$ erg s$^{-1}$. For a
distance of 8--10 kpc, this luminosity corresponds to a flux of $(0.4-1.2)
\times 10^{-10}$ erg s$^{-1}$ cm$^{-2}$. The horizontal dashed line in
Fig.~\ref{hid} indicates this limiting flux.

To investigate the emission properties in this low state, we performed a
spectral and timing analysis on the last five observations (MJD
52832--52837). Figure~\ref{propeller-spec} shows the  energy spectra of
\src\ above and below the limiting luminosity, given by Eq.~(\ref{limlum}).
At $L_X < L_{\rm min}$, the spectra are significantly softer and consistent
with a power law with $\Gamma \sim 2.1$, compared to $\Gamma \sim 1.5$ in
the accretor state.  To increase the  signal-to-noise ratio of the spectra with
$L_X < L_{\rm min}$, we obtained the average spectrum of the last five
observations ($F_x < 5\times 10^{-11}$ erg s$^{-1}$ cm$^{-2}$). The 2.5-15
keV spectrum is well described by a single power law with the best-fit
photon index  $\Gamma \sim 2.0\pm0.1$, giving $\chi^2=24.3$ for 32 degrees
of freedom. \citet{tsygankov16} also found substantial softening of the
{\it Swift}/XRT spectra  in the propeller state of the BeXBs 4U\,0115+63
and V\,0332+53. These authors found that an absorbed blackbody with
$kT\approx 0.5$ keV  described the low-state spectra well. However, given
the very narrow energy range fitted in the {\it Swift}/XRT spectrum, a
power-law distribution cannot be completely ruled out. A blackbody does not
provide a good fit of the low-flux spectra in \src\ as a result
of excess emission
above 10 keV.

Figure~\ref{propeller-prof} shows the pulse profiles at various
luminosities. X-ray pulsations are detected even below $L_{\rm min}$. The
last significant (at 99\% confidence level) detection of pulsations
occurred for observation  70096-01-24-02 (MJD 52833.2, $f_x=3.1 \times
10^{-11}$ erg s$^{-1}$ cm$^{-2}$), while for 70096-01-24-01 (MJD 52834.2,
$f_x=2.7 \times 10^{-11}$ erg s$^{-1}$ cm$^{-2}$) is only marginal at 90\%
confidence level.  Below this flux, no pulsations are detected. The decrease
in X-ray pulse fraction when the source flux drops below a certain
threshold has been interpreted as due to the propeller effect
\citep{cui97,campana01}. The average power spectrum in the low state is
shown in Fig.~\ref{propeller-psd}. Power-law noise is clearly detected. The
best-fit power-law index is $0.8\pm0.2$ and the rms variability in the
frequency range 0.01--10 Hz is $\sim40$\%.

The onset of the propeller regime does not mean that the X-ray emission is
halted. Even pulsation may be detected during quiescence
\citep{rutledge07}. Several models have been put forward to explain the
X-ray emission at low luminosities. For example, a fraction of matter may
still leak through the barrier \citep{doroshenko11}. Alternatively,
high-energy radiation may also be detected as a thermal spectrum, which
likely originates in the polar caps of the neutron star surface heated
during intermittent accretion episodes or by non-uniform cooling of the
neutron star surface after a recent outburst \citep{reig14c,tsygankov16}.
Deep crustal heating, where the crust of the neutron star cools by X-ray
emission until it reaches thermal equilibrium, can also account for the
emission at quiescence \citep{brown98,wijnands13}. In these cases, the
X-ray luminosity is expected to be $L_{\rm th} \simless 10^{33}$ erg
s$^{-1}$.   These scenarios are ruled out by the observations for \src. First, the X-ray spectrum in the low (propeller) state is not
blackbody (thermal). Second,  the luminosity is significantly higher than
$L_{\rm th}$. Finally, the power spectrum of \src,  obtained from the five
observations with $L_x < L_{\rm min}$, exhibits strong low-frequency noise
and broadband rms variability (Fig.~\ref{propeller-psd}). A flat spectrum
(white noise), that is, an absence of variability, is expected in the case of
emission from hot polar caps \citep[][and references therein]{reig14c}.

Another mechanism that can produce X-rays when the centrifugal barrier is
at work is magnetospheric accretion \citep{king94,campana95,corbet96}. In
this scenario, gravitational energy is released  by accreting material that
is deposited in the magnetosphere. X-ray pulsations may still be produced
by the azimuthal asymmetry of the rotating magnetospheric boundary. On
first approximation, the accreted luminosity will be driven by the same
mass transfer as in the accretor regime, but with the magnetospheric radius
replacing the neutron star radius $L_{m}\approx GM_{\rm NS}\dot{M}/R_m$.
The highest luminosity in this regime will occur when the magnetospheric
radius equals the co-rotation radius $R_m=R_{\rm co}=(GM_{\rm NS}P_{\rm
s}^2)^{1/3}/4\pi^2$. Hence \citep{campana02}

\begin{equation}
\label{magac}
L_{\rm min}(R_m) = 2.4 \times 10^{35} 
k^{7/2} B_{12}^2 P_{\rm s}^{-3} M_{1.4}^{-1} R_{6}^6 \, \,  {\rm erg \,
s^{-1}}
.\end{equation}

This luminosity also represents the level at which the propeller barrier
closes completely. The lowest X-ray luminosity in \src, $L_{X, \rm min}\sim
10^{35}$ erg s$^{-1}$, is still considerably higher than the
highest luminosity that can be emitted in the propeller regime as given by
Eq.~(\ref{magac}). 

Unfortunately, {\it RXTE} was not sensitive to faint observations. The
source became too faint for the sensitivity of the detectors. The
observations of \src\ stopped before the propeller barrier closed
completely. Although the source did enter the propeller accretion regime,
it did not reach a stable configuration that would allow us to distinguish
between different emission scenarios in quiescence.

\section{Conclusion}

We have performed a timing and spectral X-ray analysis of \src\ during the
2003 outburst. As the X-ray flux decreases from the peak of the outburst to
quiescence,  the source enters different accretion regimes. At the highest
flux, the X-ray luminosity of \src\ may have briefly reached the critical
luminosity, and the source may have entered the super-critical accretion
regime.  However, the absence of a diagonal branch in the
hardness-intensity diagram and the absence of a shift from anti-correlation
to correlation in the flux vs photon index diagram indicate that \src\ was
in the sub-critical state most of the time. At the end of the outburst,
when the flux decreased below $\sim 10^{-10}$  erg cm$^{-2}$ s$^{-1}$, the
source entered the propeller regime. In this state, the spectra became
softer and the X-ray pulsation vanished. Unfortunately, the source flux at
the end of the outbursts reached the sensitivity limit of the instruments
and the observations ended before a stable quiescent state was reached. The
lower-flux observations of \src\ represent a transition state between the
onset of the propeller effect and quiescence.  Based on the positive
correlation of the energy of this component with X-ray flux and pulse
phase, we tentatively associate the 10 keV feature with a cyclotron line.
The discovery of the optical counterpart and the determination of the
distance will solve the question of whether \src\ entered the
super-critical regime during the 2003 outburst.

\begin{acknowledgements}

Skinakas Observatory is a collaborative project of the University of Crete
and the Foundation for Research and Technology-Hellas. This publication
makes use of data products from the Two Micron All Sky Survey, which is a
joint project of the University of Massachusetts and the Infrared
Processing and Analysis Center/California Institute of Technology, funded
by the National Aeronautics and Space Administration and the National
Science Foundation. This work has made use of NASA's Astrophysics Data
System Bibliographic Services and of the SIMBAD database, operated at the
CDS, Strasbourg, France. This research has made use of MAXI data provided
by RIKEN, JAXA and the MAXI team.

\end{acknowledgements}

\bibliographystyle{aa}
\bibliography{../../artBex_bib}

\end{document}